\documentclass[useAMS,usenatbib]{mn2e}
\usepackage{times}
\usepackage{graphicx}
\usepackage{amssymb}
\usepackage{latexsym}
\usepackage{subfigure}
\usepackage{subeqnarray}
\usepackage{natbib}

\newcommand{\mnras}{MNRAS}
\newcommand{\apj}{ApJ}
\newcommand{\aap}{A\&A}
\newcommand{\pasj}{PASJ}

\title[On the accuracy of the Perturbative Approach for Strong Lensing]{On the accuracy of the Perturbative Approach for Strong Lensing: Local Distortion for Pseudo-Elliptical Models}
\author[H.S.~D\'umet-Montoya et al.]{Habib S. D\'umet-Montoya$^{1,2}$\thanks{E-mail: hdumetm@cbpf.br}, Gabriel~B.
Caminha$^{1,2}$, Bruno~Moraes$^{1,2}$, Martin Makler$^{1,2}$,
\newauthor Mandeep~S.~S. Gill$^{3,4}$, and Bas\'ilio X. Santiago$^{2,5}$ \\
$^{1}$Instituto de Cosmologia, Relatividade e Astrof\'isica --- ICRA, Centro
Brasileiro de Pesquisas F\'isicas\\
Rua Dr. Xavier Sigaud 150, CEP 22290-180, Rio de Janeiro, RJ, Brazil\\
$^{2}$Laborat\'orio Interinstitucional de e-Astronomia --- LIneA, \\
Rua Gal. Jos\'e Cristino 77, CEP 20921-400, Rio de Janeiro, RJ, Brazil\\
$^{3}$ Kavli Institute for Particle Astrophysics \& Cosmology, SLAC National Accelerator Laboratory \\
2575 Sand Hill Road, Menlo Park, CA 94025, USA \\
$^{4}$ Center for Cosmology and Astro-Particle Physics, Department of Physics,
The Ohio State University\\
191 West Woodruff Avenue, Columbus, OH 43210\\
$^{5}$ Departamento de Astronomia, Universidade Federal do Rio Grande do Sul\\ Av. Bento Gon\c{c}alves 9500, Porto Alegre, RS 91501-970, Brazil}

\begin{document}

\pagerange{\pageref{firstpage}--\pageref{lastpage}} \pubyear{2011}

\maketitle

\label{firstpage}

\begin{abstract}

The Perturbative Approach (PA) introduced by \citet{alard07}  provides analytic solutions for gravitational arcs by solving the lens equation linearized around the Einstein ring solution. This is a powerful method for lens inversion and simulations in that it can be used, in principle, for generic lens models. In this paper we aim to quantify the domain of validity of this method for three quantities derived from the linearized mapping: caustics, critical curves, and the deformation cross section (i.e. the arc cross section in the infinitesimal circular source approximation). We consider lens models with elliptical potentials, in particular the Singular Isothermal Elliptic Potential and Pseudo-Elliptical Navarro--Frenk--White 
models. We show that the PA is exact for this first model. For the second, we obtain constraints on the model parameter space (given by the potential ellipticity parameter $\varepsilon$ and characteristic convergence
$\kappa_s$) such that the PA is accurate for the aforementioned quantities. In this process we obtain analytic expressions for several lensing functions, which are valid for the PA in general. The determination of this domain of validity could have significant implications for the use of the PA, but it still needs to be probed with extended sources.
\end{abstract}


\section{Introduction}

Gravitational arc systems can be used as a powerful probe of the matter distribution of galaxies and galaxy clusters acting as
lenses \citep{kovner89,miralda93,hattori97}. Further, their abundance can be used to constrain cosmological models \citep{bartelmann1998,oguri01,golse2002, bartelmann03}. This motivated several arc searches to be carried out, both in wide field surveys   \citep[][Erben et al. 2012, in prep.]{Gladders03, sdss1,legacy,CASSOWARYmethod, SBAS7,CS82, RCS2,WenHanJiang2011,more11,Bayliss2012,Wiesner2012}, as well as in images targeting know clusters \citep{Luppino99,Zaritsky03,Smith2005, Sand2005, Hennawi2008,kausch2010, Horesh2010,sogras1}. 
Upcoming wide field imaging surveys, such as the Dark Energy Survey\footnote{\texttt{www.http://www.darkenergysurvey.org/}} \citep[DES;][]{annis05,des05}, which started operations in 2012, 
are expected to detect of the order of $10^3$ strong lensing systems, about an order of magnitude increase with respect to the current largest surveys.
  
Two primary approaches have been followed in practical applications of gravitational arc systems.
On the one hand, \textit{inverse  modelling} attempts to ``deproject'' the arcs in individual lens
systems to determine lens and source properties
\citep{1993A&A...273..367K,gravlens,golse2002,comerford06,lensview,
  jullo07, 2010Sci...329..924J}. On the other hand, \textit{arc  statistics}  \citep{wu93,1994ApJ...431...74G,bartelmann94}  aims at counting the number of arcs in cluster samples and comparing with the predictions from cosmological models.

Both approaches require the lens equation to be solved numerically for finite sources numerous times.
The inverse modelling typically needs arc images obtained from a multidimensional space of
source positions and lens parameters scanned during the minimization process to find the best solution for sources and lenses \citep[e.g.,][]{lensview}. For this reason analyses using the inverse modelling are often restricted to simple lens models, in particular models with elliptic lens potentials (so-called pseudo-elliptical) and/or to point sources, for example, considering bright spots in arcs as multiple images of point sources
 \citep{gravlens,lensview,jullo07,oguri10}.

For arc statistics, predictions for the arc cross section must be derived as a function of source and lens properties and the cosmological model, again by obtaining a large set of arc images
\citep{miralda93b,bartelmann94,mene01,mene03,oguri03}. The cross section is then convolved with the distribution of lens properties expected in a given cosmology and convolved with the source distribution. Another approach is to use directly high resolution N-body simulations obtaining arc images by ray-tracing through the mass distribution for a large number of sources \citep{Horesh11,DUNEsims,Boldrin2012}. 

It is therefore useful to develop approximate methods for obtaining gravitational arcs, which will be particularly useful given the increase of strong lensing systems to be discovered by the next generation wide-field surveys.
A most promising technique for this purpose is given by the Perturbative Approach \citep{alard07,alard08}, which provides an approximate solution for the lens equation close to the Einstein ring, leading to analytic solutions for arcs. 

The power of this approach is that it  can be applied, in principle, to generic lens models,  including those arising from simulations. The method is suitable for large tangential arcs, since the solutions are accurate for images located close to the Einstein ring corresponding to the circularly averaged lensing potential. 

Another important feature of the method is that it naturally reproduces arcs resulting from the merger of multiple images, which cannot be accounted for with other approximate methods for arcs proposed in the literature \citep[e.g.,][]{keet_es,fedeli05}. Such merger arcs are key for lens inversion methods and also play an important role in the arc cross section \citep{rozo08}. 

The Perturbative Approach has already been used for inverse modelling in 
\citet{alard09,alard10}. Given that it reproduces arc contours that can be associated to isophotes, it could also be used to simulate the brightness distribution of arcs, in a similar way to what was implemented in 
 \citet{paintarcs} for arc shaped contours. 

An important issue for practical applications of this approach is the determination of its domain of validity. 
This topic is discussed in \citet{alard07}, comparisons with arc simulations are presented in \citet{peirani2008}, and a recent work by \citet{japas} has investigated arcs in several configurations for a pseudo-elliptical model in this approach.  However a systematic study of its limit of applicability has not yet been carried out.
In this paper we make a first attempt to determine a domain of validity of the method in terms of the parameter space of the lens model. We will restrict to the simple case of pseudo-elliptical models, which are nevertheless widely used for the inverse modelling. Moreover, for simplicity, we will restrict the comparisons with the exact solution for three quantities connected to arcs, but which do not involve the lensing of finite sources.
We expect that the limits here obtained can be connected to the domain of validity for arcs and extended for more general models, but this is left to subsequent explorations.

In this work, our purpose is twofold. The first is to explore the
application of the Perturbative Approach to determine quantities arising from the local lens mapping, such as the 
arc cross section for infinitesimal circular sources (deformation
cross section). The second is to determine a domain of validity such
that the critical curves, caustics, and deformation cross section are accurately obtained.
This study is performed for the pseudo-elliptical Navarro--Frenk--White model (PNFW), determining regions of its parameter space where the Perturbative Approach provides a good approximation for these quantities.
We also consider the Singular Isothermal Elliptic Potential
(SIEP) model and show that the solution of the Perturbative Approach
is exact in this case.

The outline of this paper is as follows: in Sec. \ref{bstgl} we
present a few basic results of gravitational lensing theory,
introduce the radial lens models to be used in this work, and discuss models with elliptic lensing potentials.
In Sec. \ref{pert_appro_sec} we review the Perturbative
Approach, present its application to the computation of the deformation cross section, and discuss its implementation to pseudo-elliptical models.
In Sec. \ref{limits_validity} we establish a metric for the comparison
between the Perturbative Approach and the exact solution for 
critical curves and caustics and determine a domain of validity for the Perturbative
approach. In Sec. \ref{conclud} we summarize the results and present concluding
remarks.

\section{Basics of Gravitational Lensing: Definitions and Notation}
\label{bstgl}

In this section we present a brief review of the lensing theory to set up the notation and to define the quantities associated with pseudo-elliptical models. For a more detailed description see, e.g., chap. 8 of \citet{SEF}, chap. 6 of \citet{SThe} and chap. 3 of \citet{morro}.

The \textit{lens equation} relates the two-dimensional position (with respect to the optical axis)
of the observed images $\bmath{\xi}$ to those of the sources $\bmath{\eta}$. We may choose a a length scale $\xi_0$ and define $\bmath{x}=\bmath{\xi}/\xi_0$ and $\bmath{y}=\bmath{\eta}/\eta_0$,
with $\eta_0 \equiv {D_{\rm OS} \over D_{\rm OL}} \xi_0$, where $D_{\rm OL}, D_{\rm OS}$ are the angular diameter distances from the observer to the lens and source respectively. Using these definitions the lens equation is written as
\begin{equation}
\bmath{y}=\bmath{x}-\bmath{\alpha}(\bmath{x}) =
\bmath{x}-\nabla_{\bmath{x}}\varphi(\bmath{x}), \label{lens_eq_ad}
\end{equation}
where $\bmath{\alpha}(\bmath{x})$ is the ``dimensionless'' deflection angle and $\varphi(\bmath{x})$ is the ``dimensionless'' \textit{lensing potential},

The local distortion in the lens plane is described by the Jacobian matrix of
eq. (\ref{lens_eq_ad}) 
\begin{equation}\label{jacobian_matrix}
\mathbfss{J} = \left( { \partial \bmath{y} \over \partial \bmath{x}} \right)_{ij}=
\delta_{ij}-\partial_i\alpha_j(\bmath{x}).
\end{equation}
The two eigenvalues of this matrix are written as  $\lambda_r=1-\kappa+\gamma$ and
$\lambda_t=1-\kappa-\gamma$, where $\kappa$ and $\gamma$ are the 
convergence and the shear given below. Points satisfying the conditions $\lambda_{r,t}=0$ define the radial and tangential critical curves respectively. Mapping these curves onto the source plane, we obtain
the caustics. 

For axially symmetric models the deflection angle, convergence and shear are given by
\begin{eqnarray}
\alpha(x)&=&\frac{d\varphi_0(x)}{dx}=x\frac{\bar{\Sigma}(\xi_0 x)}{\Sigma_{\rm crit}}, \label{angle_axial}\\
\kappa(x)&=&\frac{1}{2}\left[ \frac{\alpha(x)}{x}+\frac{d \alpha(x)}{d  x}\right], \label{conv_axial} \\
\gamma(x)&=&\frac{1}{2}\left[ \frac{\alpha(x)}{x}-\frac{d\alpha(x)}{d  x}\right].\label{shear_axial}
\end{eqnarray}
where $\bar{\Sigma}(\xi_0 x)$ is the mean surface density within a radius $x$ and $\Sigma_{\rm crit}$ is the critical surface density.

In this work, one model we will make use of is the Singular Isothermal Sphere (SIS), which is useful to model lenses at the galactic scale. Its dimensionless lensing potential, deflection angle, convergence and shear are given by \citep{turner84,SEF, 2009MNRAS.398..607V} 
\begin{equation}
\varphi_0(x)=x, \ \ \alpha(x)=1, \ \ \kappa(x)=\gamma(x)=\frac{1}{2}, \label{pot_sis}
\end{equation}
where we choose the Einstein Radius to be the characteristic scale 
\[
\xi_0=R_{\rm E}={ \sigma^2_v \over G\Sigma_{\rm crit}},
\]
where $\sigma_v$ is the velocity
dispersion.  From this potential analytic solutions of the lens equation can be obtained for finite sources \citep{2006MNRAS.365.1243D,dm_thesis}.

We will also make use of the Navarro--Frenk--White model \citep[][hereafter NFW]{nfw96,nfw97}, often used to represent lenses in the galaxy to galaxy cluster mass scales. This model has two independent parameters $r_s$ and $\rho_s$. 
By fixing $\xi_0=r_s$ and defining the characteristic convergence as
\begin{equation}
\kappa_s =  \frac{\rho_s r_s }{\Sigma_{\mbox{\tiny crit}}},
\end{equation}
the lensing potential is given by \citep{bartelmann96}
\begin{equation}
\varphi_0(x)=4\kappa_s\left(\frac{1}{2}\log^2{\frac{x}{2}}-2\,{\rm
arctanh}^2{\sqrt{\frac{1-x}{1+x}}} \right),\label{pot_nfw}
\end{equation}
which is a function of the parameter $\kappa_s$ alone.

Models with elliptical potentials (the so-called pseudo-elliptical models) provide simple analytical solutions for some lensing quantities \citep{1987ApJ...321..658B,kassiola93,kneib01}. They have been widely used in lens inversion problems 
and are implemented in several public codes for lens inversion such as {\it Gravlens} \citep{gravlens}, {\it Lensview}
\citep{lensview}, {\it Lenstool} \citep{jullo07}, and {\it glafic} \citep{oguri10}.
They have also been used for arc simulations \citep{oguri02,mene03,mene07}. 

Pseudo-elliptical models, with potential $\varphi_\varepsilon(\bmath{x})$, are built from a given axially symmetric potential $\varphi_0(x)$ by replacing the radial coordinate $x$  by
\begin{equation}
 \tilde{x}=\sqrt{a_{1}\,x^2_1+a_{2}\,x^2_2} = x \Delta_\phi,
\label{subti-ellip}
\end{equation}
where
\begin{equation}
\Delta_\phi \equiv \sqrt{a_{1}\,\cos^2{\phi}+a_{2}\,\sin^2{\phi}},
\label{Deltaphi}
\end{equation}
such that the  ellipticity of the lensing potential is 
\[
\varepsilon_\varphi=1-\sqrt{\frac{a_1}{a_2}}, 
\]  
where the orientation was chosen such that the major axis of the ellipse is along the $x_1$ axis (i.e., $a_2>a_1$).
The deflection angle, convergence, and shear can be written as combinations of the lensing functions of the corresponding axially symmetric model for any choice of $a_1$ and $a_2$ \citep{dm_2011}. 

The SIEP and PNFW models are obtained by following this procedure for the potentials given in eqs. (\ref{pot_sis}) and (\ref{pot_nfw}), respectively.

\section{Perturbative Approach}
\label{pert_appro_sec}

For a given lens model, the Perturbative Approach allows one to obtain
analytic solutions for arcs as perturbations of the Einstein Ring solution.
In this work we investigate the limits of applicability of the Perturbative Approach, by considering simple non-axially symmetric models and by looking at local properties of the lens mapping, instead of lensed finite sources.

In this section we  briefly review the Perturbative Approach and use it for the derivation of the caustics and critical curves, the deformation cross section and quantities needed for its computation. The method is also applied to models with elliptical lensing potentials.

\subsection{Lens Equation}

The gist of the Perturbative Approach for gravitational arcs developed
by \citet{alard07,alard08} is to obtain an analytic solution for the
lens equation considering the lens as a perturbation of an axially
symmetric configuration and the source position as a small deviation
from the optical axis (i.e., positioned transversely away from perfect
observer--lens--source alignment). In other words, the arcs are found
as perturbations of the Einstein Ring configuration. In this work we will consider the thin lens and the single lens plane approximations, which imply a unique solution for the Einstein Ring \citep{2008MNRAS.391..668W}.

The Einstein Ring is the image of a source aligned with an axially symmetric lens (with lensing potential $\varphi_0$). Its radius $x_{\rm E}$ is obtained by solving the $\lambda_t(\bmath{x})=0$ at the centre of the source plane,  i.e.
\begin{equation}
x-\frac{d\varphi_0}{d x}=0.
\end{equation}
Arcs can be obtained by perturbing the equation above either by shifting the position of the source away from the optical 
axis and/or by adding a non-circular perturbation to the lensing potential. 
These perturbations are described by
\begin{equation}\label{pert_approx_1}
\bmath{y}= \delta y, \ \
\varphi(\bmath{x})=\varphi_0(x)+ \delta \psi(\bmath{x}).
\end{equation}
These perturbations are assumed to be of the same order in $\epsilon$ (the strength of the perturbation) throughout the following calculations, such that 
\begin{displaymath}
\delta y = \epsilon \overline{y}, \ \ \delta \psi(\bmath{x})= \epsilon \overline{\varphi}(\bmath{x}). 
\end{displaymath}
The response to such perturbations is given by the displacement of the radial coordinate in the lens plane\footnote{Note that in
\citet{alard07,alard08} $\xi_0=x_{\rm E}$ was used as a characteristic 
scale. This choice is equivalent to setting $x_{\rm E}=1$ in our equations. In this 
work we have made the choice of keeping $x_{\rm E}$ explicitly in the 
equations for more generality, allowing us, for example, to choose another characteristic scale of the problem.}, i.e., $x = x_{\rm E}\rightarrow  x=x_{\rm E} + \delta x$ where we also assume the same order in $\epsilon$ such that $\delta x= \epsilon\overline{x}$.

To find $\epsilon \overline{x}$ we solve eq.~(\ref{lens_eq_ad}) by expanding the solution 
around $x=x_{\rm E}$. Expanding the lensing potential in a Taylor series 
around $x=x_{\rm E}$, we have
\begin{equation}
\varphi(\bmath{x}) =\sum_{n=0}^{\infty}\left[C_n+f_n(\phi)\right](\epsilon\overline{x})^n, \label{pot_expand}
\end{equation}
where
\begin{eqnarray}
C_n & \equiv& \frac{1}{n!}\left[\frac{d^n\varphi_0}{d\,x^n}\right]_{x=x_{\rm E}}, \nonumber \\
\label{fn-cn_def} \\
f_n(\phi) &\equiv& \frac{1}{n!}\left[\frac{\partial^n\delta \psi}{\partial\,x^n}
\right]_{x=x_{\rm E}}=\frac{\epsilon}{n!}\left[\frac{\partial^n\overline{\varphi}}{\partial\,x^n}
\right]_{x=x_{\rm E}}. \nonumber
\end{eqnarray}

Inserting $x=x_{\rm E} + \epsilon \overline{x}$ and (\ref{pot_expand}) into
eq.~(\ref{lens_eq_ad}), we find that the  resulting equation at zeroth order in $\epsilon$ is 
\begin{equation}
x_{\rm E}=C_1,
\end{equation}
which is the Einstein Ring equation. Using the relations above and $\delta x= \epsilon \overline{x}$,  the resulting equation at the first order in $\epsilon$ is given by 
\begin{eqnarray}
y_1=(\kappa_2 \delta x -f_1)\cos{\phi}+\frac{1}{x_{\rm E}}\frac{d f_0}{d
\phi}\sin{\phi},   \nonumber  \\
\label{lens_eq_pa_1} \\
y_2=(\kappa_2 \delta x -f_1)\sin{\phi}-\frac{1}{x_{\rm E}}\frac{d f_0}{d
\phi}\cos{\phi}, \nonumber
\end{eqnarray}
where $\kappa_2 \equiv 1-2C_2$. From eqs. (\ref{angle_axial})--(\ref{shear_axial}) 
we have \begin{displaymath}
\frac{d^2\varphi_0(x)}{d\,x^2}=2\kappa(x)- \frac{\alpha(x)}{x},
\end{displaymath} and therefore $\kappa_2$ can be expressed as 
\begin{equation} 
\kappa_2=2-2\kappa(x_{\rm E}). \label{k_2-def}
\end{equation}

Eq.~(\ref{lens_eq_pa_1}) is the lens equation in the Perturbative Approach. It 
can be solved for  $\delta x$ for each angular position $\phi$ of the source, given a perturbation described by $f_n(\phi)$.
To obtain the images of a finite source, we must first parametrize its boundary. Then, by varying $\phi$ from $0$ to $2\pi$, each point of that boundary is mapped to the lens plane  through eqs. (\ref{lens_eq_pa_1}). As a result, a new equation with separated radial and angular components is formed, whose solution is obtained straightforwardly \citep{alard07,alard08,peirani2008,dm_thesis} 

It is important emphasize that the solutions ($x_i,\phi_i$) of eq.~(\ref{lens_eq_pa_1}) are valid only to first order in the perturbations in eq.~(\ref{pert_approx_1}), i.e. only for points near the Einstein Ring. For points far from this curve, the solutions are not expected to be highly accurate. 
For this reason, the Perturbative Approach is particularly useful for applications involving tangential arcs. 
In this work, instead of using finite sources, we focus  on the potential applicability of this method to quantities based on the local mapping as a first step to quantify the differences with the exact solutions.

\subsection{Local Mapping}
The Jacobian matrix for the lens mapping is
\begin{equation}
\mathbfss{J} = \left(\frac{\partial \bmath{y}}{\partial
\bmath{x}}\right)_{ij}=\sum_{k}\left(\mathbfss{J}_{S\rightarrow
L,pol}\right)_{ik}\,\left(\mathbfss{J}_{L,pol\rightarrow
cart}\right)_{kj},\label{jacob_matrix_pa}
\end{equation}
where $\mathbfss{J}_{S\rightarrow L,pol}$ is the Jacobian of the transformation 
from the lens plane to the source plane in polar coordinates from eq. (\ref{lens_eq_pa_1})
and $\mathbfss{J}_{L,pol\rightarrow cart}$ is the standard Jacobian matrix 
from polar to Cartesian coordinates. The calculation of the eigenvalues of the 
lens mapping is then straightforward from the equation above and they are given by 

\begin{equation}\label{eigv_1-2_pa}
\begin{array}{l}
\lambda_{t}=-\frac{1}{x}\left[\frac{1}{x_{\rm E}}\frac{d^2f_0}{d\phi^2}-(\kappa_2 \delta x -f_1)\right], \\
\lambda_{r}=\kappa_2.
\end{array}
\end{equation}
Therefore, the radial coordinate of the tangential critical curve is 
\begin{equation}
x_{t}(\phi)=x_{\rm E}+ \delta x_t(\phi)=x_{\rm
E}+\frac{1}{\kappa_2}\left(f_1+\frac{1}{x_{\rm
E}}\frac{d^2f_0}{d\phi^2}\right),\label{xcc_pm}
\end{equation}
and the parametric equations of the critical curve are simply
\[
\begin{array}{l}
x_{1t}=x_t(\phi)\cos{\phi}, \\ 
x_{2t}=x_t(\phi)\sin{\phi}.
\end{array}
\]
Inserting $\delta x_t$ in eq.~(\ref{lens_eq_pa_1}), the parametric equations of the tangential 
caustic are found to be
\begin{eqnarray}\label{ca_pm}
y_{1t}= \frac{1}{x_{\rm E}}\frac{d^2f_0}{d\phi^2}\cos{\phi}+\frac{1}{x_{\rm
E}}\frac{df_0}{d\phi}\sin{\phi},\\
y_{2t}= \frac{1}{x_{\rm E}}\frac{d^2f_0}{d\phi^2}\sin{\phi}-\frac{1}{x_{\rm
E}}\frac{df_0}{d\phi}\cos{\phi}. \nonumber
\end{eqnarray}

\subsection{Constant Distortion Curves}
\label{cd_curves}
For infinitesimal circular sources, the length-to-width ratio of arcs can
be approximated by the ratio of the eigenvalues of the lens mapping Jacobian 
matrix \citep{wu93,bartelmann94,haman97}
\begin{equation}
\frac{L}{W}\simeq |R_\lambda(\bmath{x})| \equiv
\left|\frac{\lambda_{r}(\bmath{x})}{\lambda_{t}(\bmath{x})}\right|\label{lw}.
\end{equation}

Under this approximation, it is possible to define a region where 
gravitational arcs are expected to form by fixing a value for the threshold length to width ratio $R_{\rm th}$. 
Such region is limited by the curves $R_\lambda =\pm R_{\rm th}$ (constant 
distortion curves). Although the condition (\ref{lw}) does not hold for merger 
arcs \citep{rozo08}, nor for large or elliptical sources, the curves 
defined above still provide a typical scale for the region of arc formation. 
In this work, we adopt the common choice $R_{\rm th}=10$ (unless explicitly 
stated otherwise). We denote the radial coordinates of these curves as 
$x_\lambda$. They are obtained by solving $R_\lambda(\bmath{x})=\pm R_{\rm th}$, 
with $\lambda_r$ and $\lambda_t$ given in the Perturbative Approach by eq.~(\ref{eigv_1-2_pa}). It follows 
that
\begin{equation}\label{x_l-mper}
x_{\lambda}(\phi)=x_{t}(\phi)\times\left\{\begin{array}{cc} \frac{R_{\rm th}}{R_{\rm th}-1},
& R_\lambda=+R_{\rm th},\\ & \\
\frac{R_{\rm th}}{R_{\rm th}+1}, & R_\lambda=-R_{\rm th}.\end{array}\right.
\end{equation}

The constant distortion curves in the lens plane are therefore self-similar to the 
tangential critical curve. The mapping of these curves to the source plane is 
done by substituting $\delta x_\lambda=x_\lambda-x_{\rm E}$ in eq. 
(\ref{lens_eq_pa_1}). For instance, the curve $R_\lambda=+R_{\rm th}$ has the 
following parametric equations
\begin{equation} \label{x_l_src_mper}
\begin{array}{l}
y_{1\lambda}=\frac{f_1+\kappa_2 x_{\rm E}}{R_{\rm
th}-1}\cos{\phi}+\frac{1}{x_{\rm E}}\left[ \left(\frac{R_{\rm th}}{R_{\rm
th}-1}\right)\frac{d^2f_0}{d\phi}\cos{\phi}+\frac{d
f_0}{d\phi}\sin{\phi}\right],\\
 \\
y_{2\lambda}=\frac{f_1+\kappa_2 x_{\rm E} }{R_{\rm
th}-1}\sin{\phi}+\frac{1}{x_{\rm E}}\left[ \left(\frac{R_{\rm th}}{R_{\rm
th}-1}\right)\frac{d^2f_0}{d\phi}\sin{\phi}-\frac{d
f_0}{d\phi}\cos{\phi}\right].
\end{array}
\end{equation}

The parametric equations of the $R_\lambda=-R_{\rm th}$ curve are given by the
expressions above with the substitution 
$R_{\rm th}-1 \rightarrow R_{\rm th}+1$. There is no self-similarity between 
these curves and the tangential caustics.

\subsection{Deformation Cross Section}
\label{d_cross_section}

As mentioned in the introduction, the \textit{arc cross section} is a key ingredient in arc statistics calculations. 
Is is defined as the effective area in the source 
plane such that sources within it will be mapped into images with 
$L/W \geq R_{\rm th}$. The definition of this area must take into account the 
image multiplicity given the source position \citep[i.e. multiply-imaged 
regions are counted multiple times, see e.g.,][]{mene03}. The computation of 
the arc cross section in general demands ray-tracing simulations, which 
are computationally expensive 
\citep{miralda93b,bartelmann94,mene01,mene03,oguri03}. An alternative 
is to use the infinitesimal circular source approximation, 
eq. (\ref{lw}), which allows the computation of the arc cross section to be carried out directly 
from the local mapping from lens to source plane. In this case, 
$\sigma_{R_{\rm th}}$ is computed in the lens plane by 
\begin{equation}
\sigma_{R_{\rm th}} = \eta^2_0\tilde{\sigma}_{R_{\rm th}} =
\eta^2_0\int_{|R_\lambda|\geq R_{\rm th}}\frac{d^2\,x}{|\mu(\bmath{x})|}
\label{dcs_dim}
\end{equation}
\citep[see, e.g.,][]{fedeli05,dm_2011,caminha09},
where $\mu=\left(\lambda_r\lambda_t\right)^{-1}$ is the magnification
and the integral is performed over the region of arc formation above the chosen
threshold. The quantity $\tilde{\sigma}_{R_{\rm th}}$ is known as the
dimensionless deformation cross section. 

In the Perturbative Approach, the magnification 
can be written from eq. (\ref{eigv_1-2_pa}) as
\begin{equation}
\left|\mu(\bmath{x})\right|^{-1}=
\frac{\kappa^2_2}{x}\left\{\begin{array}{lc}x_t(\phi)-x, & x < x_t(\phi),\\ \\
x-x_t(\phi), & x > x_t(\phi), \end{array} \right.
\end{equation}
where $x_t(\phi)$ is given in eq. (\ref{xcc_pm}). Inserting the equation above
in eq.~(\ref{dcs_dim}) and integrating the radial coordinate within the lower and upper limits given in eq. (\ref{x_l-mper}), it is straightforward to obtain
\begin{equation}
\tilde{\sigma}_{R_{\rm th}}=\kappa_2^2\frac{|R_{\rm th}|^2+1}{\left(|R_{\rm
th}|^2-1\right)^2}\int_0^{2\pi} x^2_t(\phi)\,{\rm
d}\phi\label{sec-choq-pert-app2}.
\end{equation}

Note that $\tilde{\sigma}_{R_{\rm th}} \propto R^{-2}_{\rm th}$ for $R_{\rm th}
\gg 1$, as expected from the behaviour of the deformation cross section with
$R_{\rm th}$ \citep{rozo08,caminha09}.

For axially symmetric models ($x_t=x_{\rm E}$) the cross section is given simply by
\begin{equation}
\tilde{\sigma}_{R_{\rm th}}=2\pi\kappa_2^2x^2_{\rm E}\frac{R_{\rm
th}^2+1}{\left(R_{\rm th}^2-1\right)^2}.\label{sigma_mp-axial}
\end{equation}

The expression above  is exact for the SIS model \citep{1995A&A...297....1B}. For other axially symmetric models this expression is still an approximation, since the curves $R_\lambda=\pm R_{\rm th}$ are obtained approximatively. 

\begin{figure*}
\begin{center}
\resizebox{\hsize}{!}{
\subfigure{\includegraphics{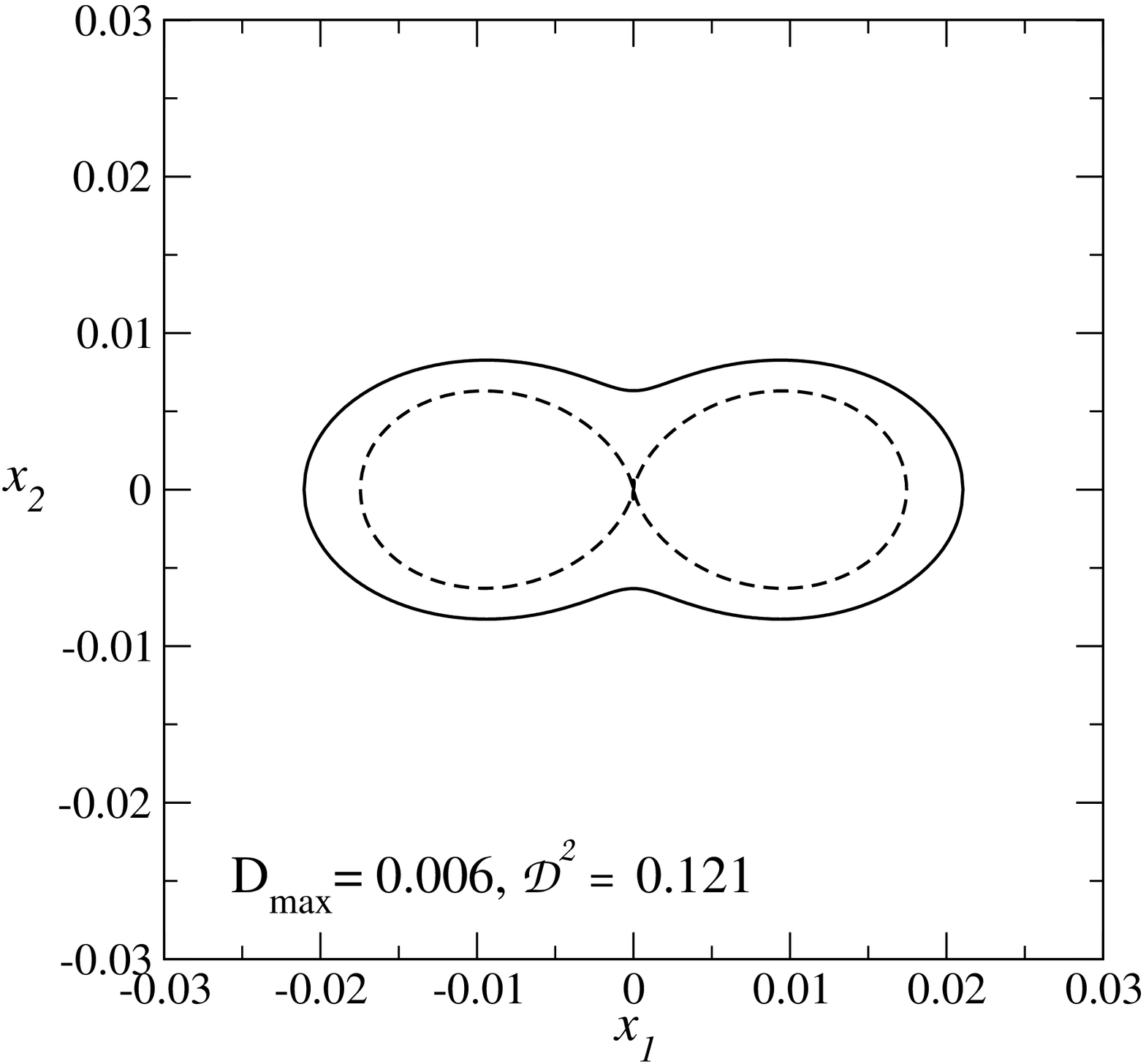}} \hspace{0.5cm}
\subfigure{\includegraphics{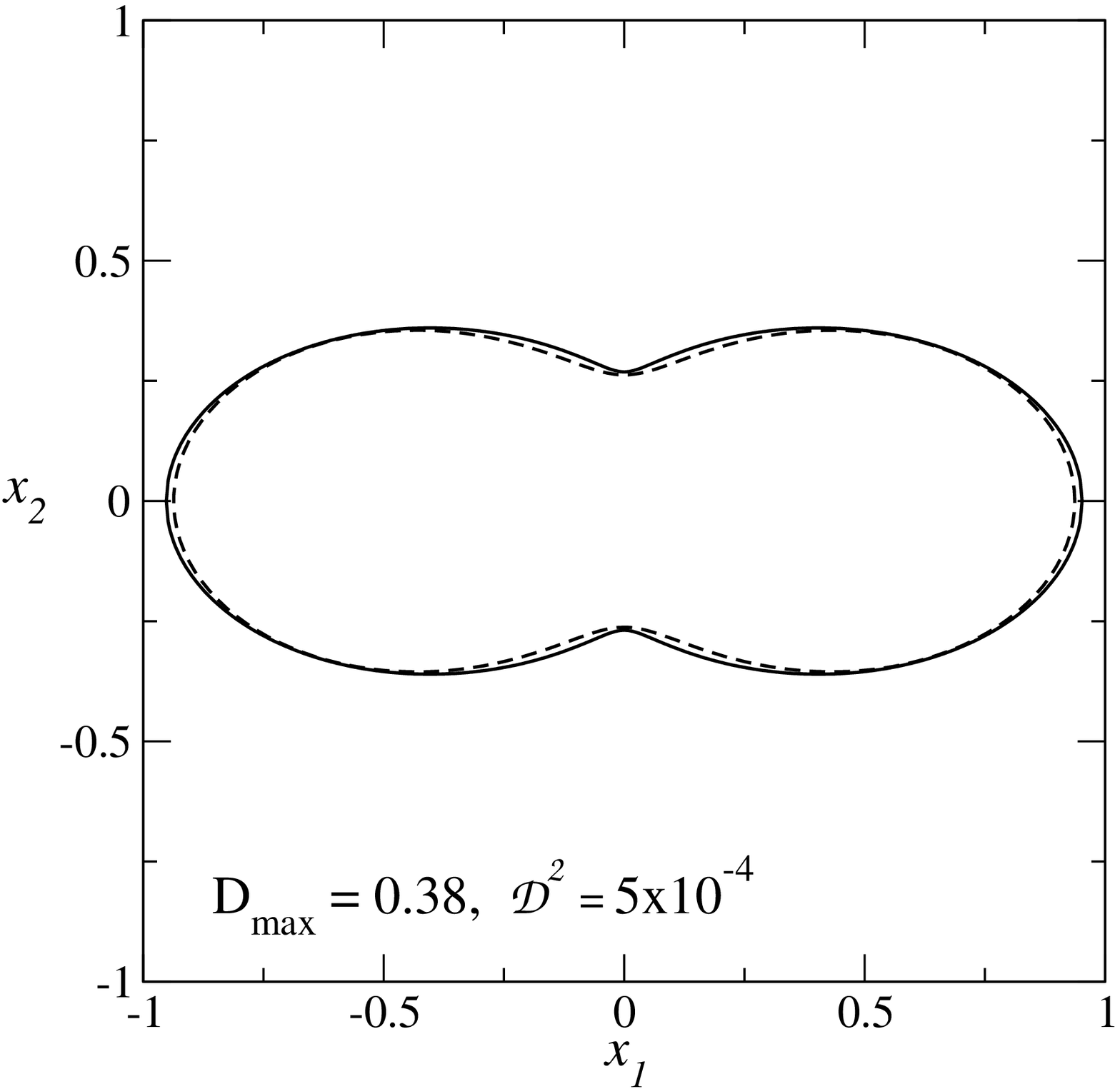}} \hspace{0.5cm}
\subfigure{\includegraphics{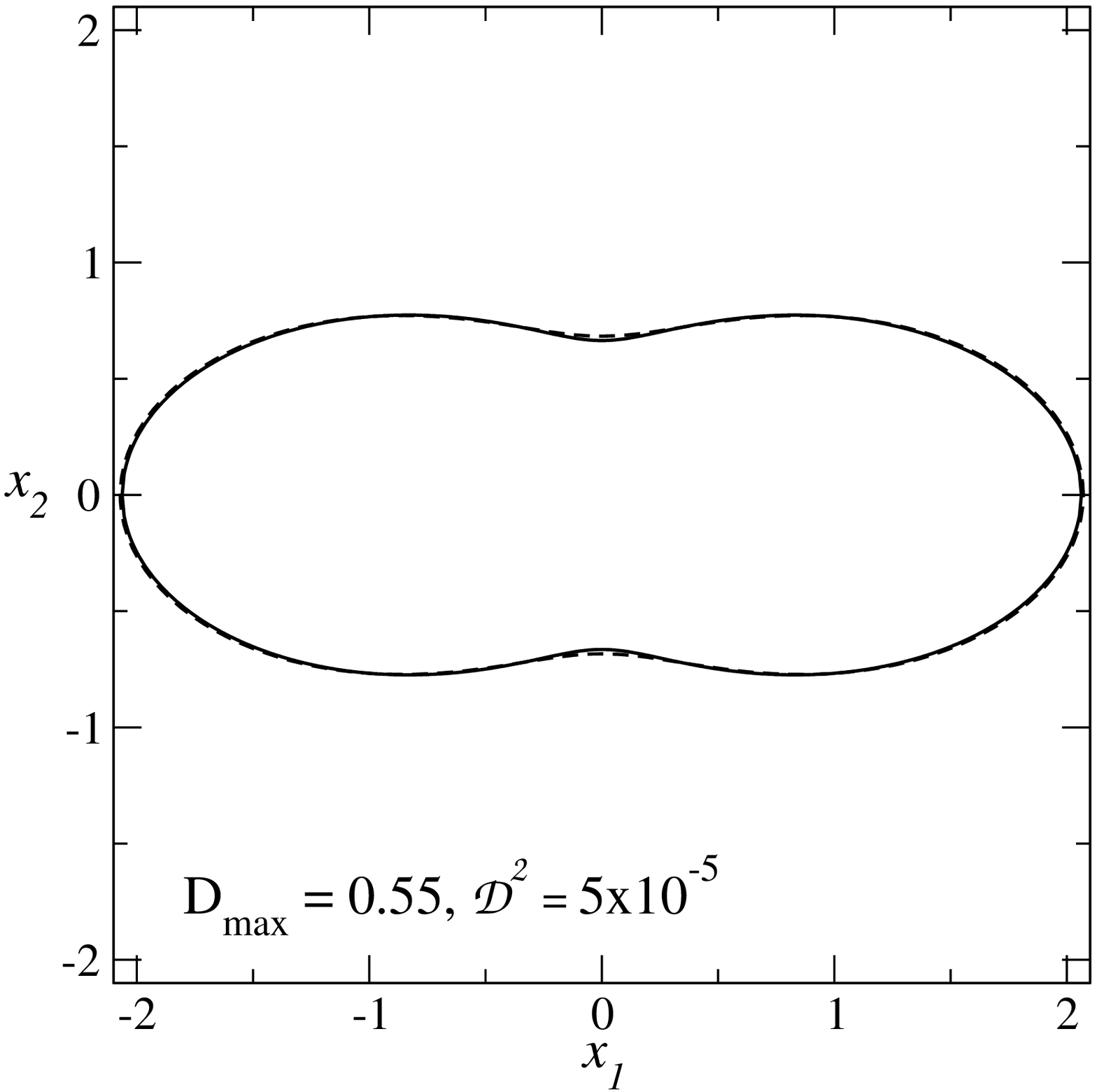}}}
\vspace*{0.5cm}
\resizebox{\hsize}{!}{
\subfigure{\includegraphics{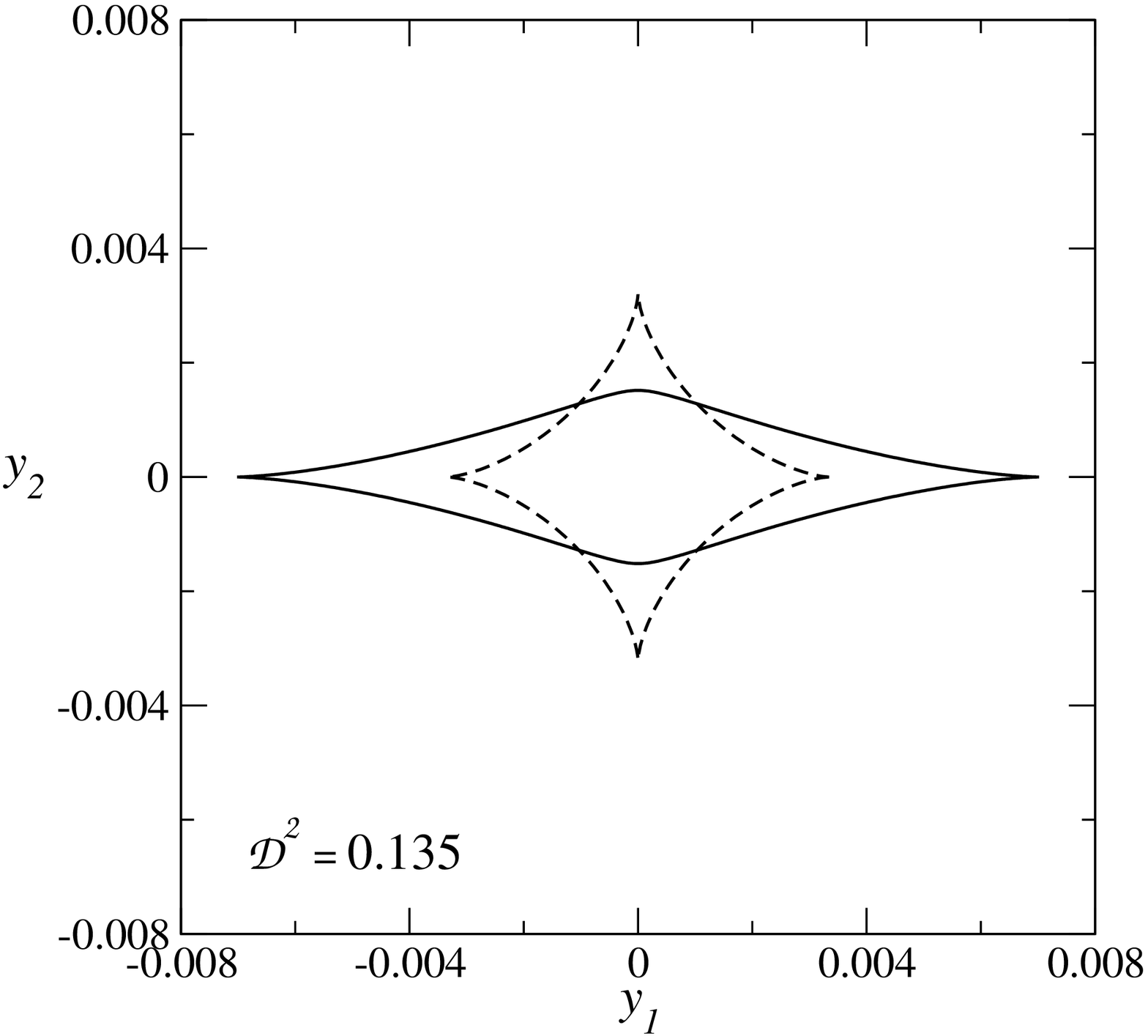}} \hspace{0.5cm}
\subfigure{\includegraphics{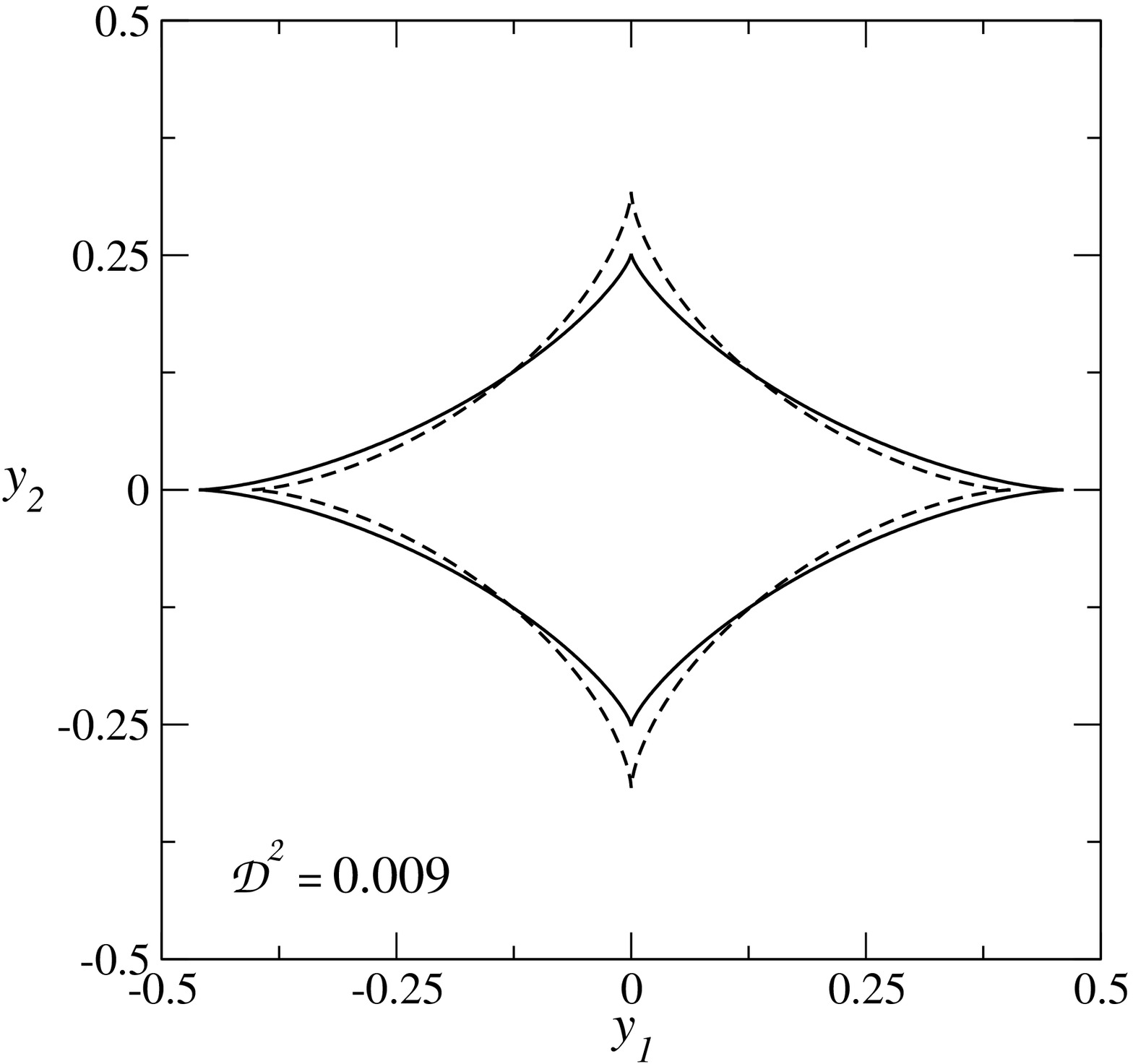}} \hspace{0.5cm}
\subfigure{\includegraphics{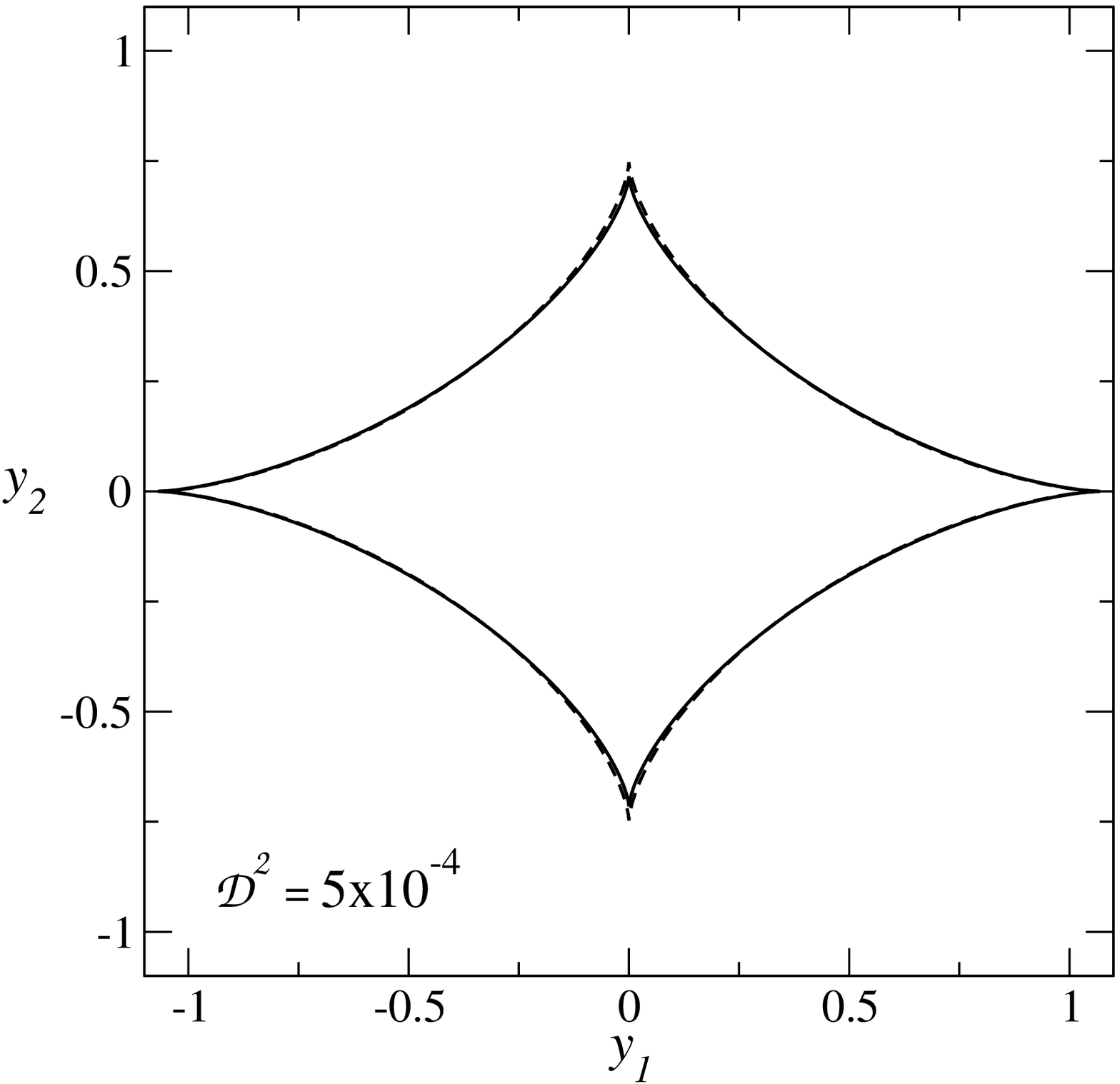}}}
\caption{\label{pnfw_ccurves_mp-se} Critical curves (top panels) and caustics
(bottom panels) obtained with the Perturbative Approach (dashed lines) and the
exact solution (solid lines) for the PNFW model: $\kappa_s=0.1$ and
$\varepsilon=0.2$ (left), $\kappa_s=0.5$ and $\varepsilon=0.32$ (middle) and
$\kappa_s=1.0$ and $\varepsilon=0.35$ (right). The values of $\varepsilon$ in
the middle and right panels were chosen by imposing $\mathcal{D}^2=5\times
10^{-4}$ for critical curves and caustics, respectively (see Sec.
\ref{limits_validity}).}
\end{center}
\end{figure*}

\subsection{Perturbative Functions for Pseudo-Elliptical Models}
We write the elliptical potential as
\begin{equation}
\varphi_\varepsilon(\bmath{x})=\varphi_0(x)+\left[\varphi_0(\tilde{x})-\varphi_0(x)\right],
\end{equation}
such that the perturbed potential becomes
\[\delta \psi(x,\phi)=\varphi_0(\tilde{x})-\varphi_0(x). \] 
From the definitions (\ref{fn-cn_def}) and using the identities
(\ref{angle_axial})--(\ref{shear_axial}), it follows that
\begin{eqnarray}
f_1 &= &\frac{\tilde{x}_{\rm E}}{x_{\rm E}}\alpha(\tilde{x}_{\rm
E})-\alpha(x_{\rm E}), \nonumber \\
\frac{d f_0}{d\phi}&=& \frac{x^2_{\rm E}}{2\tilde{x}_{\rm E}}
\alpha(\tilde{x}_{\rm E})(a_{2}-a_{1})\sin{2\phi}, \label{dfn_pe}\\
\frac{d^2 f_0}{d\phi^2}&=& \frac{x^2_{\rm E}}{\tilde{x}_{\rm E}}
\alpha(\tilde{x}_{\rm E})(a_{2}-a_{1})\cos{2\phi} \nonumber \\
& &
- \frac{\gamma(\tilde{x}_{\rm E})}{2}\left[ \frac{ x^2_{\rm E}}{\tilde{x}_{\rm
E}}(a_{2}-a_{1})\sin{2\phi}\right]^2, \nonumber
\end{eqnarray}
where $\alpha$ and $\gamma$ are the deflection angle and shear of the 
corresponding axially symmetric lens. These expressions hold for any
parametrization of the lensing potential ellipticity and for any
pseudo-elliptical lens \citep{dm_2011}.

For small values of the lensing potential ellipticity, eqs. (\ref{dfn_pe})
reduce to
\begin{eqnarray}
f_1 &=&  \frac{a_1-a_2}{2}\kappa(x_{\rm E})x_{\rm E}\cos{2\phi}\, + \mathcal{O}(\epsilon^2),
\nonumber\\ 
\frac{df_0}{d\phi}&=& \frac{a_2-a_1}{2} x^2_{\rm E}\sin{2\phi}\, + \mathcal{O}(\epsilon^2),
\label{fn_pe_small}\\
\frac{d^2f_0}{d\phi^2}&=& (a_2-a_1)x^2_{\rm E}\cos{2\phi}\, + \mathcal{O}(\epsilon^2).\nonumber
\end{eqnarray}
From eq. (\ref{xcc_pm}) and the expressions above, we have
\begin{equation} x_t(\phi)=x_{\rm E}\left[1+
\frac{a_2-a_1}{2}\left(\frac{2-\kappa(x_{\rm
E})}{\kappa_2}\right)\cos{2\phi}\right],\label{xt_approx}\end{equation}
and inserting this into eq (\ref{sec-choq-pert-app2}) we get
\begin{equation}\label{cross_section-approx}
\tilde{\sigma}_{R_{\rm th}}= 2\pi x^2_{\rm E}\frac{R_{\rm th}^2+1}{\left(R_{\rm
th}^2-1\right)^2}
\left[\kappa^2_2+\frac{1}{8}\left(1+\frac{\kappa_2}{2}\right)^2(a_2-a_1)^2
\right].
\end{equation}
Thus, for small ellipticities, the deviation with respect to the axially symmetric case is quadratic.

Instead of using $a_2$ and $a_1$ it is more intuitive to express the results in terms of the ellipticity of the potential. Several parameterizations have been used to define the ellipticity in this context.
From now on, we adopt the 
convention \citep{1987ApJ...321..658B,gk02, dm_2011} 
\begin{equation}
a_{1}=1-\varepsilon, \quad a_{2}=1+\varepsilon, \label{gk_par}
\end{equation}
where $\varepsilon$ is the potential ellipticity parameter. The connection to the ellipticity of the mass distribution $\varepsilon_\Sigma$ depends on the model. For the SIEP $\varepsilon_\Sigma = 3 \,\varepsilon$ to first order in $\varepsilon$ \citep{kassiola93}. For the PNFW model this relation depends on $\kappa_s$ and expressions for 
$\varepsilon_\Sigma(\varepsilon,\kappa_s)$ are provided in \citet{dm_2011}.

Fig. \ref{pnfw_ccurves_mp-se} 
shows the comparison for caustics and critical curves between the Perturbative Approach and the exact solution 
for the PNFW model for different values of $\kappa_s$ and $\varepsilon$.

\subsection{Singular Isothermal Elliptic Potential}\label{SIEP}

One of the simplest and most often used lens models is given by the SIEP. For this model, using expressions  (\ref{pot_sis}) in Eq. (\ref{dfn_pe}), the perturbative functions are
\begin{eqnarray}
f_1 &=& \Delta_{\phi}-1, \nonumber \\
\frac{df_0}{d\phi}&=&(a_2-a_1)\frac{\sin{2\phi}}{2\Delta_{\phi}}, \label{pert_funct_siep}\\
\frac{d^2f_0}{d\phi^2}&=& a_1\,a_2 \Delta^{-3}_{\phi}- \Delta_{\phi}, \nonumber
\end{eqnarray}
where $\Delta_{\phi}$ is given in eq. (\ref{Deltaphi}).  When substituted into eqs. (\ref{lens_eq_pa_1}) the expressions above lead to
\begin{equation}\label{lens_eq_siep}
y_1=x\left(1-\frac{a_1}{\tilde{x}}\right)\cos{\phi} \ \
{\rm and} \; \;
y_2=x\left(1-\frac{a_2}{\tilde{x}}\right)\sin{\phi},
\end{equation}
which are the components of the lens equation of this model without any approximation. Hence, the solution of the Perturbative Approach is exact in the case of lensing by the SIEP model. 
 
The same conclusion does not hold for the PNFW model. We will thus investigate the domain of validity for this model in the next section.

\section{Limits of validity of the Perturbative Approach for the PNFW model}
\label{limits_validity}

\begin{figure*}
\begin{center}
\resizebox{\hsize}{!}{
\subfigure{\includegraphics{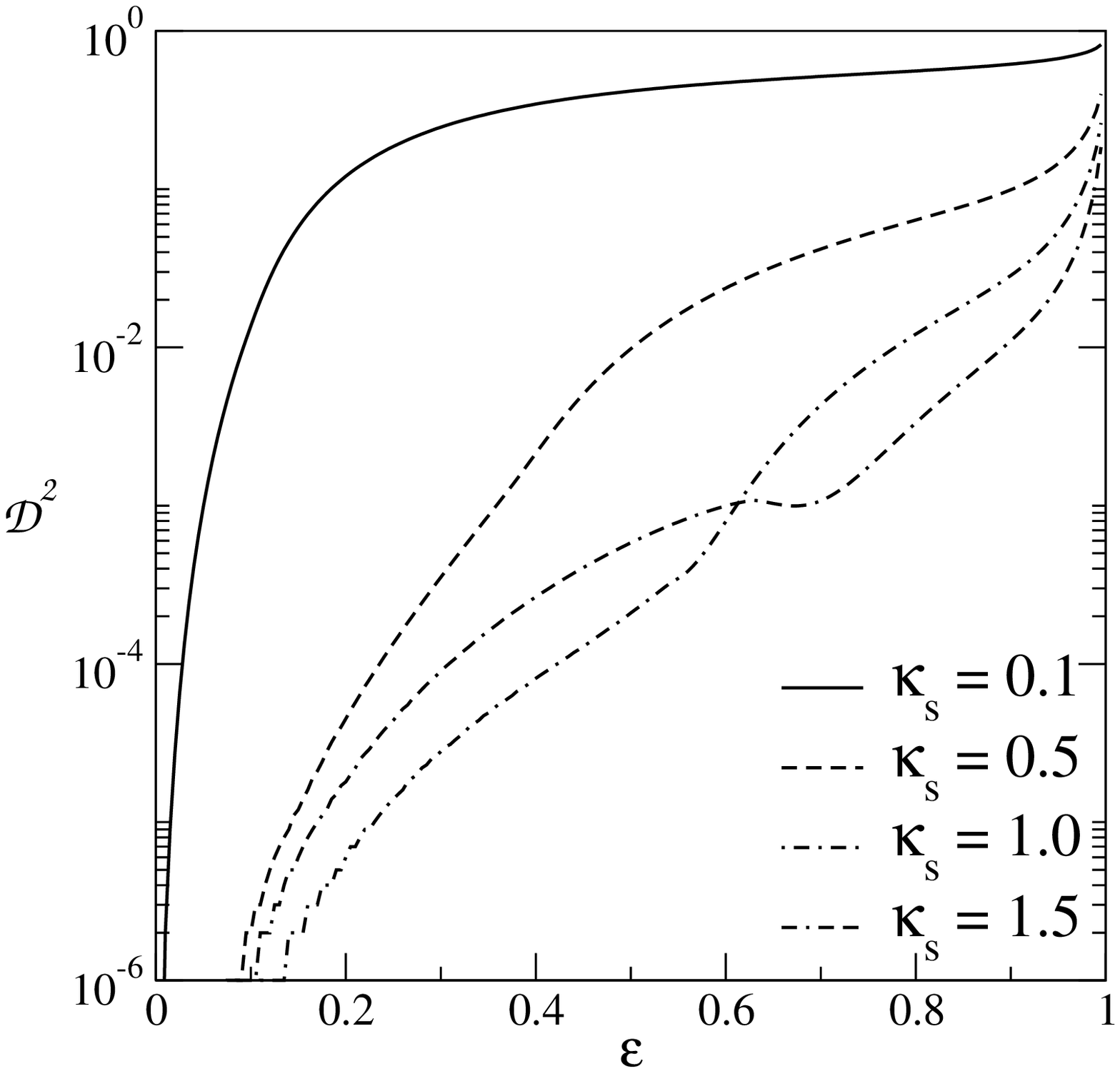}} \hspace{1.5cm}
\subfigure{\includegraphics{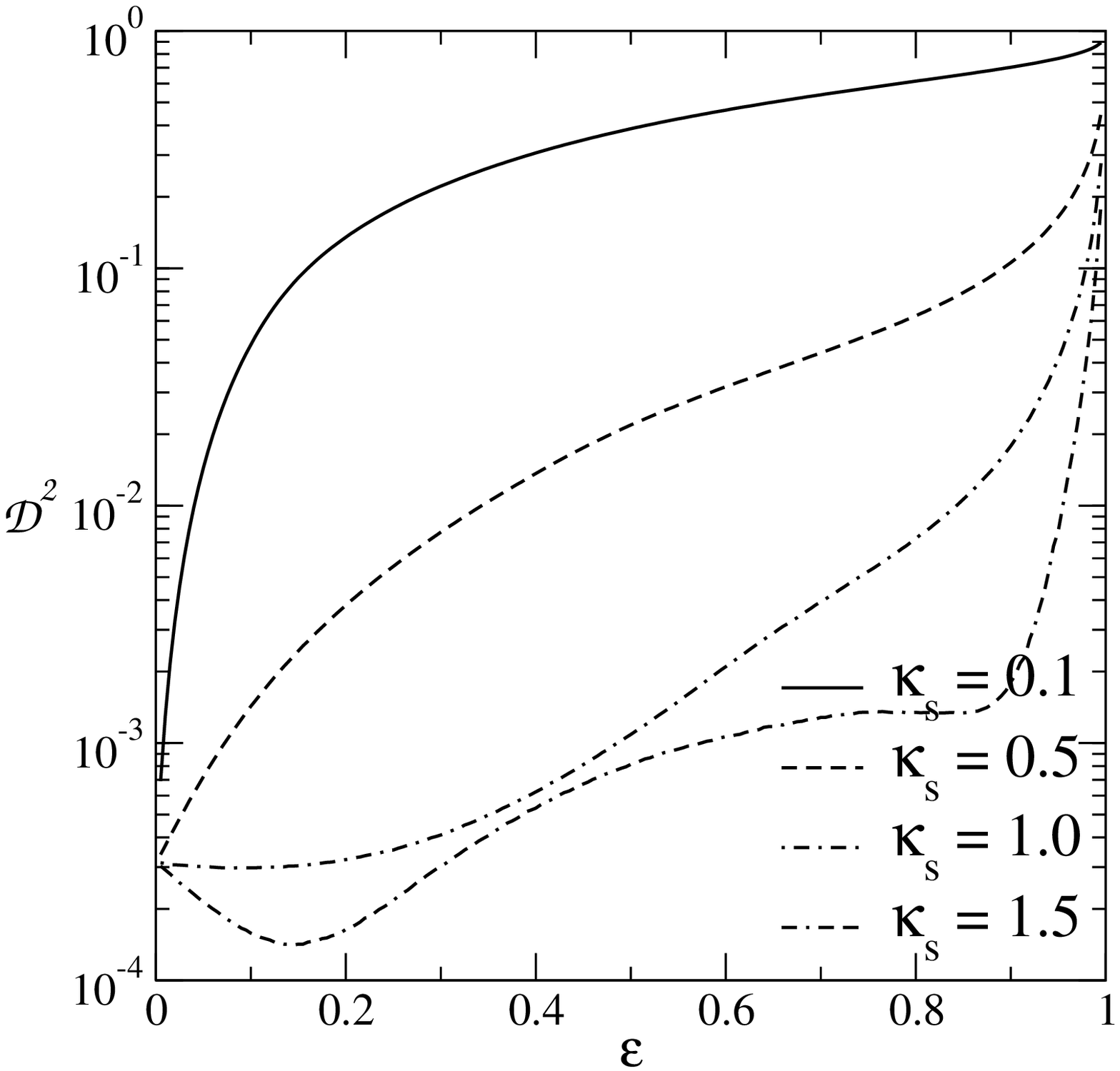}}}
\caption{\label{d2_pnfw} Mean weighted squared radial fractional difference
$\mathcal{D}^2$  as a function of the ellipticity parameter for the PNFW model for various values of the characteristic convergence $\kappa_s$. Left
Panel: For Critical Curves. Right Panel: For Caustics.
}
\end{center}
\end{figure*}

Previous attempts to quantify the differences between exact and
perturbative solutions were carried out in the literature.
\citet{alard07}
proposed a method based on the relative importance of the third-order
term in the Taylor series of the gravitational potential. 
\citet{japas} performed a qualitative analysis of a particular arc configuration, varying some
of the system parameters and establishing criteria based on the position and multiplicity of the images.
However, they did not define a metric to compare the solutions nor carry on the analysis for 
more general configurations.

Investigating the domain of validity of the Perturbative Approach with finite sources would
require a large parameter space to be probed, including the lens and source parameters and their relative positions. On the other hand, as a starting point, we may look at quantities that are dependent only on the lens, such as the tangential caustic and critical curve and the deformation cross section (the latter will depend also on the choice of $R_{\rm th}$). Besides reducing the parameter space --- for example, for $\varepsilon$ and $\kappa_s$ in the PNFW case --- it is simpler to define metrics to quantify the deviation of the perturbed solution from the exact one. We expect that the constraints on the domain of validity determined from the quantities above can be connected to those arising from the images of finite sources. 
Thus, exploring the simplest case before may provide guidance to the determination of the domain of validity of the method finite sources in the future. 
Setting a domain of validity from the lens model alone may provide a rapid method to adjudicate
validity of the Perturbative Method {\it a priori}, just from the lensing potential,
without the need of obtaining images of the sources.

In this section, we shall attempt to quantify the deviation of critical curves 
and caustics using a figure-of-merit akin to the one proposed in \cite{dm_2011}. We will then 
compare the deformation cross sections and, finally, combine the results to obtain limits that define a 
region in the parameter space of PNFW models where the Perturbative Approach 
can be used to accurately obtain local properties of a given lens system.

\subsection{Limits for critical curves and caustics}
\label{sec-curvelimits}

To quantify  the deviation of the solution of the Perturbative Approach from the exact one for critical curves and caustics we use a figure-of-merit defined as the mean weighted squared fractional radial difference between the curves, i.e.\footnote{Expression (\ref{D2_pm}) is formally equal to the one proposed in \citet{dm_2011}, where it was used
to compare an isocontour of $\kappa$ to an ellipse. Here the same expression is used to compare two solutions for caustics or critical curves.}
\begin{equation}
\mathcal{D}^ 2  \equiv \frac{\sum_{i=1}^N\,w_i[x_{\rm ES}(\phi_{i})-x_{\rm
PA}(\phi_{i}) ]^2 }%
  {\sum_{i=1}^N w_i\, x_{\rm ES}^2(\phi_{i})},
  \label{D2_pm}
 \end{equation}
where $x_{\rm ES}(\phi_{i})$ and $x_{\rm PA}(\phi_{i})$ are the radial
coordinates of the tangential curves (either critical curves or caustics) obtained
from the exact solution and with the Perturbative Approach, respectively.  These are computed on a discrete set of $N$ points defined by the polar angle $\phi_i$. Further, $w_i \equiv \phi_i-\phi_{i-1}$ is a weight to account for a possible non-uniform distribution of points in $\phi$.  

Choosing a cut-off value for $\mathcal{D}^2$, we can define a range in $\varepsilon$ for which the curves obtained with both the exact and perturbative solutions will be similar enough to each other. 
The cut-off value is then chosen by visually comparing the exact and perturbative solutions for the critical curves and caustics associated with several values of $\mathcal{D}^2$, for combinations of the PNFW lens parameters.

Before presenting the results, we should stress a technical point. 
In the particular case of caustics, calculating the two functions in the same 
polar angle becomes a non-trivial issue.  This is because 
in general, the source 
plane points ($y_{1t},y_{2t}$) are not equally distributed in angle, as they are 
obtained scanning angular values in the lens plane which map nonlinearly to angular values in the source plane.  Thus in general, a source plane 
angle does not correspond to the same lens plane angle. Yet, to compute 
$\mathcal{D}^2$ for the caustics, it is necessary that both $x_{\rm ES}$ and $x_{\rm PA}$ be calculated at the same polar angle position in the source plane. Thus, to enforce this last point, we first determine the polar angle 
corresponding to each point ($y_{1t},y_{2t}$) obtained with the exact solution, 
i.e., \[\phi_S=\arctan{\left(\frac{y_{2t}}{y_{1t}}\right)}\] and obtain the 
corresponding radial coordinate $x_{\rm ES}=y_t(\phi_S)=\sqrt{y^2_{1t}+
y^2_{2t}}$. In the same way, we compute the polar angle of the tangential 
caustic obtained with the Perturbative Approach (which we denote by $\phi_S^{\rm 
PA}$), i.e.,\[\phi^{\rm PA}_S=\arctan{\left(\frac{y_{2t}(\phi_L)}{y_{1t}(
\phi_L)}\right)},\] where $y_{1t}$ and $y_{2t}$ are given in eq. (\ref{ca_pm}) 
and $y^{\rm PA}_t=\sqrt{y^2_{1t}+y^2_{2t}}$. We then vary the angle $\phi_L$ 
(only the interval $0\leq \phi_L \leq \pi/2$ is needed, for symmetry reasons) such that for each radial position $y_t$, 
the angles $\phi_S$ and $\phi^{\rm PA}_S$ are chosen to have the same value at step $i$. Finally, having 
determined ($y_t,\phi_S$) for the exact solution and $(y^{\rm PA}_{t},\phi^{\rm 
PA}_S)$ we proceed to compute $\mathcal{D}^2$ as in eq. (\ref{D2_pm}).

Fig. \ref{d2_pnfw} shows $\mathcal{D}^2$ as a function of
$\varepsilon$ for some values of\footnote{Throughout this work, following \citet{dm_2011}, we will consider the range $\kappa_s \leq 1.5$.} $\kappa_s$. In the left panel, the results for critical curves are shown. Since the perturbation increases with $\varepsilon$, $\mathcal{D}^2$ also increases with $\varepsilon$, as we might expect. In addition, $\mathcal{D}^2$
decreases as $\kappa_s$ increases, at least for $\kappa_s<1.0$. In the right panel, we show the results for caustics. The behaviour of $\mathcal{D}^2$ is qualitatively similar to that of critical curves, except for at the highest $\kappa_s$, where the behaviours are reversed. However, the values of $\mathcal{D}^2$ computed for caustics
are higher than the corresponding ones for critical curves, for a
given $(\kappa_s,\varepsilon)$. This means that imposing cut-off
values of $\mathcal{D}^2$ for matching caustics, we will match the corresponding critical curves automatically.
We found  by visual inspection that for $\mathcal{D}^2 \leq 5\times 10^{-4}$ there is a 
very good match for the caustic curves. In Fig. \ref{pnfw_ccurves_mp-se} we show the 
values of $\mathcal{D}^2$ calculated for each example, demonstrating visually the validity of this 
diagnostic measure. In particular, we have checked that cut-off values of 
$\mathcal{D}^2$ higher but close to our chosen limit of $5\times 10^{-4}$ are not suited for 
matching caustic curves well. 

To estimate the validity of the Perturbative Approach, \citet{alard07}
introduced the parameter $D \equiv 3|C_3|(\delta {x}_{\rm arc})^2$, where $\delta {x}_{\rm arc}$ corresponds to 
the difference between the \textit{arc contours} obtained in the perturbative approach and the Einstein radius, and $C_3$ is the third-order term in the Taylor expansion of the gravitational potential (see eq. \ref{pot_expand}). In order for the 
Perturbative Approach to be accurate, $D$ should be small.
For models based on the SIS profile, 
this condition is always true, since $C_3=0$ (which is consistent with the fact that the Perturbative Method is exact in this case). For other pseudo-elliptical 
models, usually $C_3\neq 0$. 

Here we adapt the definition of $D$ to be used for critical curves, such that $\delta x$ is now the radial deviation of these curves with respect to $x_E$.
We associate a unique value of $D$ to the tangential critical curve, using its 
maximum value over this curve, which corresponds to
\begin{equation}\label{dmax_pnfw}
D_{\rm max}=3|C_3|{\rm max}\{(x_t(\phi)-x_{\rm E})^2\},
\end{equation}
where $x_t$ is given in eq. (\ref{xcc_pm}) and $0 \leq \phi \leq 2\pi$.
Following Alard's criterion (i.e. $D_{\rm max} \ll 1$), it would be expected that 
the critical curves and caustics obtained with the Perturbative Approach would 
be close to the ones obtained in the exact case when both $\varepsilon$ and 
$\kappa_s$ are small. We compute $D_{\rm max}$ for 
the curves shown in Fig. \ref{pnfw_ccurves_mp-se}, obtaining 
$D_{\rm max}= 0.006$, $0.38$ and $0.55$ from left to right panels. Contrary to 
expectations, when $D_{\rm max}$ increases, the curves obtained with the 
Perturbative Approach become more similar to the exact solutions. 
Therefore, the criterion $D_{\rm max} \ll 1$ does not reflect the validity of 
the Perturbative Approach for these cases. Moreover, $D_{\rm max}$ is not 
scale-invariant (i.e. $D_{\rm max} \propto r^2_s$, where $r_s$ is the 
length scale of the PNFW model). These considerations show that this measure is 
not well-suited to assess the limit of validity of the method for caustics and critical curves.
This result emphasizes the relevance of our definition of $\mathcal{D}^2$ as a measure for the validity 
of the Perturbative Approach for critical curves and caustics.

\begin{figure}
\begin{center}
\resizebox{\hsize}{!}{\includegraphics{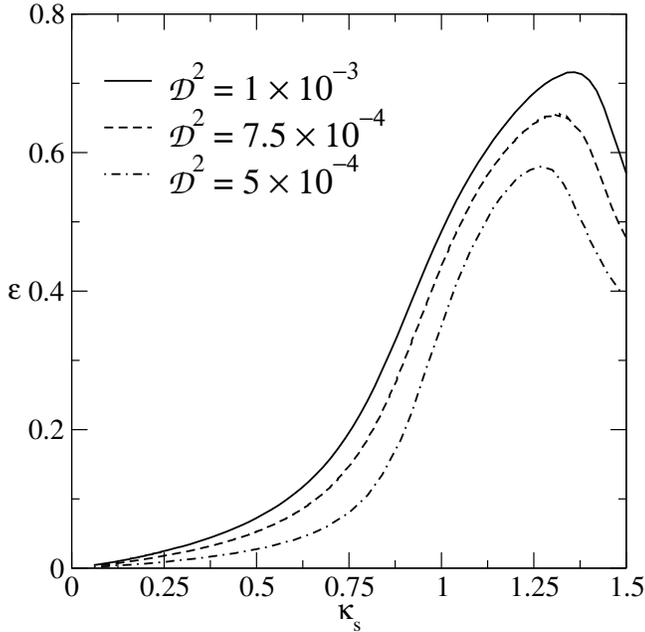}}
\caption{\label{emax_d2_pnfw} Maximum values of $\varepsilon$ obtained from some
cut-off values of $\mathcal{D}^2$ on caustics for the PNFW model.} 
\end{center}
\end{figure}

For the application of our criterion, we define $\varepsilon^{\rm 
PA}_{\rm max}$, for a given $\kappa_s$, as the ellipticity threshold giving $\mathcal{D}^2 = 5\times 10^{-4}$. This will be 
used as a measure of the limit of applicability for the Perturbative Approach for critical curves and caustics. 
Fig. \ref{emax_d2_pnfw} shows the maximum values of $\varepsilon$ as a function of $\kappa_s$ for the PNFW model, for some 
cut-off values of $\mathcal{D}^2$. 
The $\varepsilon^{\rm PA}_{\rm max}(\kappa_s)$ function shown in this figure is 
well-fitted by a Pad\'{e} approximant of the form
\begin{equation}
 \varepsilon^{\rm PA}_{\rm max}=\frac{\sum_{n=0}^4
a_n{(\kappa_s)}^n}{\sum_{m=0}^2 b_m{(\kappa_s)}^m}, \label{emax_pm_fit}
\end{equation}
with $a_n=\{-0.018, 0.235, -0.415, 0.565,-0.264\}$ and $b_n=\{2.243,-3.709,1.725
\}$.

\begin{figure*}
\begin{center}
\resizebox{\hsize}{!}{
\subfigure{\includegraphics{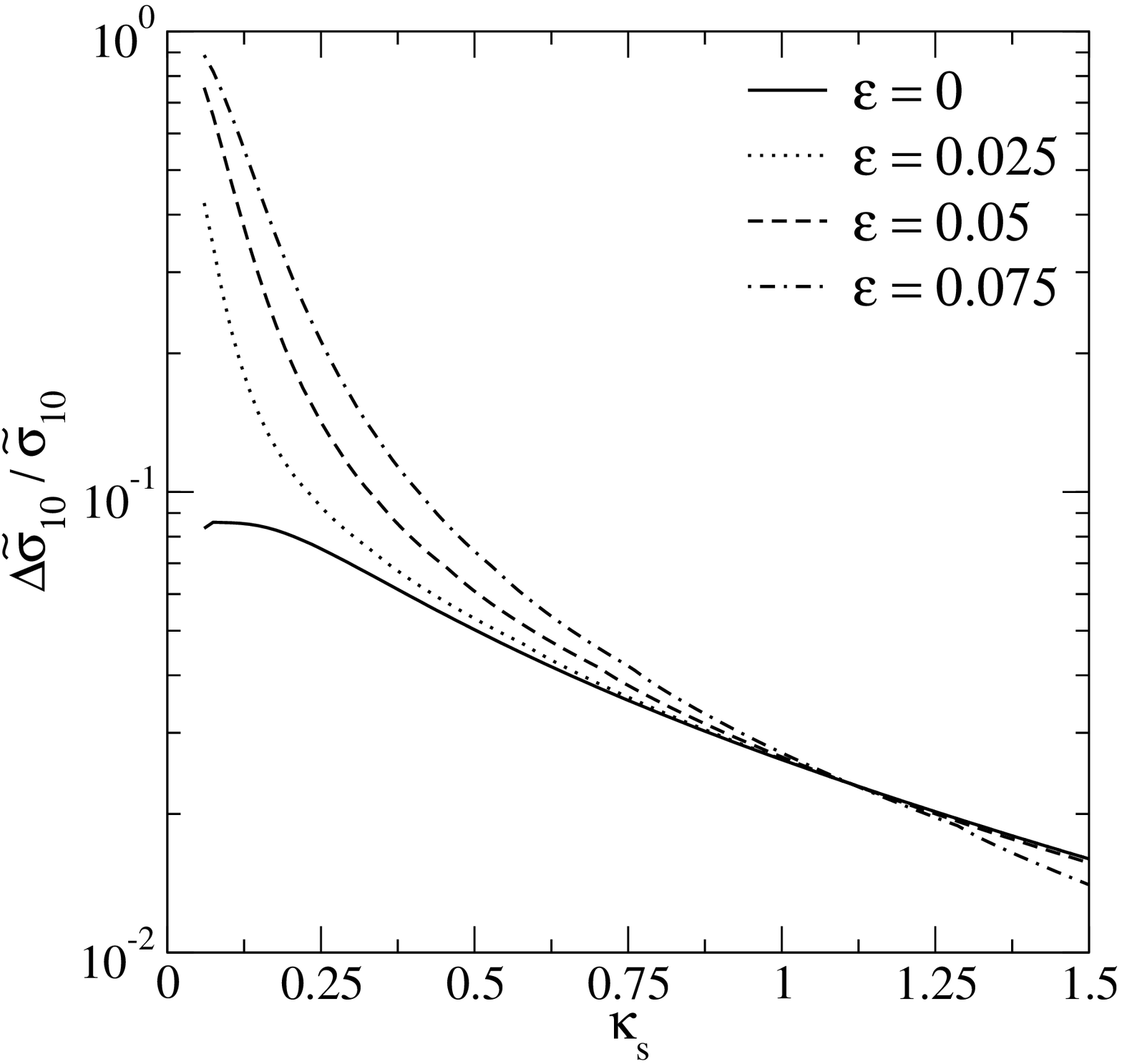}} \hspace{1.5cm}
\subfigure{\includegraphics{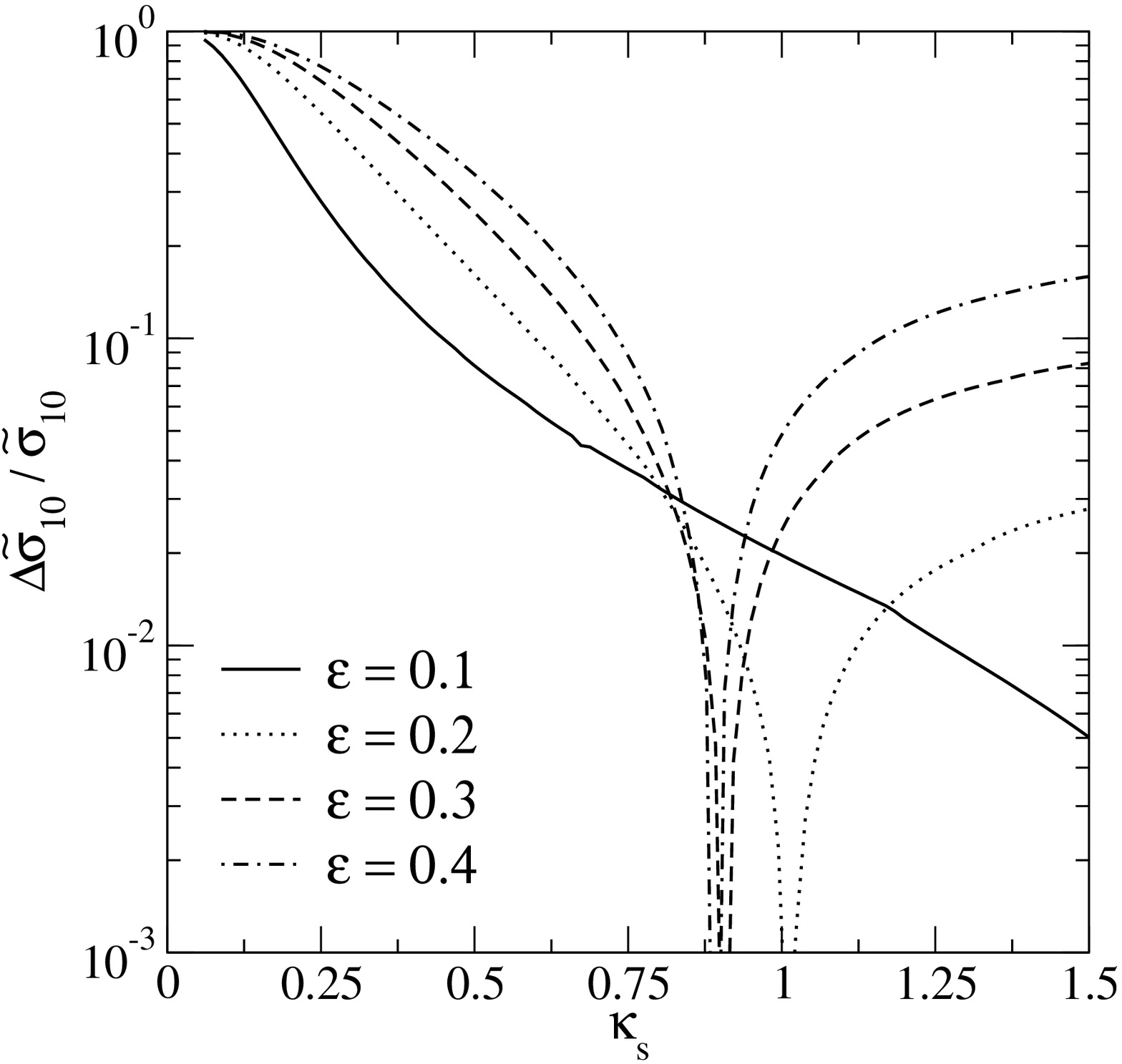}}}
\caption{\label{dcs_pnfw-e-ks} Relative deviation among deformation cross
sections for the PNFW model as a function of $\kappa_s$ for some values of
$\varepsilon$. Left panel: $\Delta \tilde{\sigma}_{10}/\tilde{\sigma}_{10}$
between exact solution and eq. (\ref{cross_section-approx}). Right panel:
$\Delta \tilde{\sigma}_{10}/\tilde{\sigma}_{10}$ between exact solution and eq.
(\ref{sec-choq-pert-app2}).}
\end{center}
\end{figure*}

\subsection{Comparison between Deformation Cross Sections}

In this section, we compare the exact and perturbative solutions for the deformation cross section in order to establish 
limits of validity for the approximation of this quantity. We then contrast these limits to those obtained for caustics and critical curves as done in Sec.~\ref{sec-curvelimits} (i.e. by imposing $\varepsilon <\varepsilon^{\rm PA}_{\rm max}$ for each $\kappa_s$). If within this regime the Perturbative Approach and the exact solution of the deformation cross section do not agree well, this can impose additional limits to the applicability of the Perturbative Approach. 

To quantify the difference between the deformation
cross sections, we compute their relative difference
\begin{equation}
\frac{\Delta \tilde{\sigma}_{R_{\rm th}}}{\tilde{\sigma}_{R_{\rm
th}}}=\left|\frac{\tilde{\sigma}_{\rm ES,R_{\rm th}}-\tilde{\sigma}_{\rm
PA,R_{\rm th}}}{\tilde{\sigma}_{\rm ES,R_{\rm th}}}
\right|,\label{dif_rela_sigma}
\end{equation}
where the subscripts ${\rm  ES}$ and ${\rm  PA}$ refer to the exact and
perturbative calculations, respectively.

In Fig. \ref{dcs_pnfw-e-ks} we show $\Delta
\tilde{\sigma}_{10}/\tilde{\sigma}_{10}$ as a function of $\kappa_s$ for some
values of $\varepsilon$. 
In the left panel we compare the exact solution with the expansion for low ellipticities in the Perturbative Approach, eq.~(\ref{cross_section-approx}), while in the right panel we compare with the general expression, eq.~(\ref{sec-choq-pert-app2})
\footnote{It should be noted that, due to the quadratic scaling with respect to $\epsilon$
of the deviation from the axially symmetric case, eq. (\ref{cross_section-approx}) is an excellent approximation for eq. (\ref{sec-choq-pert-app2}) for low ellipticities. Thus the left panel of  Fig. \ref{dcs_pnfw-e-ks} would remain essentially unchanged if we used eq. (\ref{sec-choq-pert-app2}) instead of eq.~(\ref{cross_section-approx}) there.}. 
The perturbative calculation for the axially symmetric NFW model ($\varepsilon=0$, eq. \ref{sigma_mp-axial}) is a good approximation in this case, since  $\Delta \tilde{\sigma}_{10}/\tilde{\sigma}_{10} < 10\%$ for the 
entire allowed range of $\kappa_s$. 
For values of $\varepsilon < 0.1$ the Perturbative Approach is a good approximation only for $\kappa_s \gtrsim 0.5$.
As $\varepsilon$ increases, the difference is larger at smaller values of $\kappa_s$. However, the perturbative 
calculation is accurate to within about $10\%$  for $\kappa_s \gtrsim 0.7$ up to 
$\varepsilon =0.3$ (see the right panel of Fig. \ref{dcs_pnfw-e-ks} ). 

Additionally, we computed $\Delta\tilde{\sigma}_{\rm th}/\tilde{\sigma}_{\rm th}$ as a function of the
threshold $R_{\rm th}$. We find that $\Delta\tilde{\sigma}_{\rm th}/\tilde{\sigma}_{\rm th}$ can exceed 50\% at values of $R_{\rm th} \lesssim 2.5$, since for these values of $R_{\rm th}$, the \textit{constant distortion curves} are far from the tangential critical curves, meaning that the premises of the Perturbative Approach do not apply. However, as 
$R_{\rm th}$ increases, the relative deviations among the deformation cross sections decrease. In particular, we found that for $\kappa_s \gtrsim 0.9$ and $R_{\rm th}>7.5$, these relative deviations do not depend on $R_{\rm th}$. 

\begin{figure}
\begin{center}
\resizebox{\hsize}{!}{\includegraphics{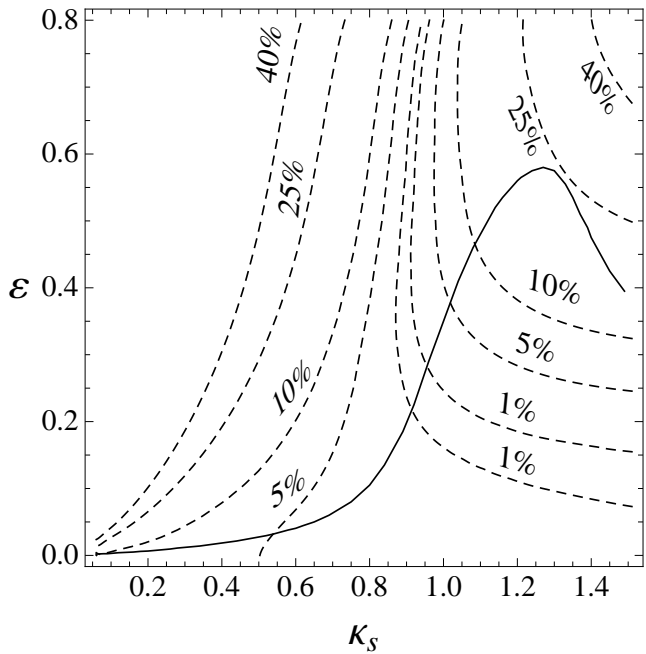}} 
\caption{\label{dcs_pnfw-space} Comparison between deformation cross sections
for the PNFW model space parameter. Contours of constant $\Delta\tilde{\sigma}_{10}/\tilde{\sigma}_{10}$.
The solid line shows the $\varepsilon^{\rm PA}_{\rm max} (\kappa_s)$ curve.} 
\end{center}
\end{figure}

In Fig.~\ref{dcs_pnfw-space} we show isocontours of $\Delta\tilde{\sigma}_{10}/\tilde{\sigma}_{10}$, for the exact and perturbative calculations, in the parameter space $\kappa_s$--$\varepsilon$ together with the curve $\varepsilon_{\max}^{\rm PA}(\kappa_s)$. We see that the constraints imposed by $\Delta\tilde{\sigma}_{10}/\tilde{\sigma}_{10}$ and 
$\varepsilon^{\rm PA}_{\rm max}$ are complementary, meaning that for $\kappa_s \lesssim 1.0$ the constraint obtained with caustics and critical curves is the strongest, while the opposite is true for $\kappa_s > 1.0$ if we impose that the maximum fractional deviation for the cross section is 10\%.

We may then combine the constraints to define a region limited approximately 
by the curves
\begin{equation}\label{region_lt-10}
\varepsilon = \left\{\begin{array}{c l} \varepsilon^{\rm PA}_{\rm
max}(\kappa_s), &  \kappa_s \lesssim 1.0,\\ 0.33, & \kappa_s>1.0. \end{array}
\right.
\end{equation}
Within this region the Perturbative Approach can replace the exact
computation of critical curves, caustics, and deformation cross section
with high accuracy.  

\section{Concluding Remarks} 
\label{conclud}

The Perturbative Approach \citep{alard07,alard08,alard09,alard10} provides analytical solutions for gravitational arcs by solving the lens equation linearised around the Einstein Ring solution. This method has a wide range of potential applications, from the inverse problem in strong lensing to fast arc simulations.
This technique goes beyond other analytical approximations in the literature in that it may be used for generic lens models (including mass distributions arising from N-body simulations) and for finite sources. 

A key aspect for practical applications of the method that has not been systematically addressed before
is the determination of its limit of validity.
Motivated by this issue, in this paper we aimed to determine the accuracy of the Perturbative Approach for 
caustics and critical curves and for the deformation arc cross section. Although these quantities do not involve 
arcs (i.e. the lensing of finite sources) they allow one to obtain limits on the accuracy of the linearized mapping from the Perturbative Approach. 
Also, the parameter space to be probed is significantly decreased, since these quantities depend basically on the lens properties and not the source ones. 

We have considered a restricted set of lens models, more specifically those with elliptical lens potentials, and in particular the PNFW and SIEP models, which are nevertheless widely used in strong lensing applications, specially for the inverse modelling. Whenever possible, we sought to derive analytic expressions for the quantities involved in the calculations, many of which are new. Some are valid for the Perturbative Approach in general, others apply to pseudo-elliptic lens models. The main results of the paper are summarized below.

We obtained analytic expressions for the constant distortion curves in the Perturbative Approach (eqs.~\ref{x_l-mper}
and  \ref{x_l_src_mper}), which, in the lens plane, are found to be self-similar to the tangential critical curve. We derived an analytic formula for the deformation cross section (eq.~\ref{sec-choq-pert-app2}), which reproduces the scaling of the arc cross section  with $R_{\rm th}$ obtained numerically in previous works. 
For axially symmetric models the cross section is obtained in closed form  (eq.~\ref{sigma_mp-axial}).

We have obtained  simple analytic expressions for the perturbative functions for pseudo-elliptical
models, which are valid for  any choice of the
ellipticity parametrization (eq. \ref{dfn_pe}). 
These expressions generalize those given in \citet{alard07,alard08}
and in \citet{japas}. 

We derive approximate solutions to the tangential critical curve (eq. \ref{xt_approx}) and
for the deformation cross section (eq. \ref{cross_section-approx}) for low ellipticities in pseudo-elliptical models. We show that the deviation of the cross section with respect to the axially symmetric case is quadratic in the ellipticity.

We have considered the SIEP and the PNFW models to represent lenses at galaxy and
galaxy cluster mass scales. We have shown that the Perturbative Approach provides the exact
solution for the SIEP model. For the PNFW model, we compared the
critical curves and caustics obtained with this approach with those
obtained with the exact solution for a wide range of values of
$\kappa_s$ and $\varepsilon$. 

We show that the criterion $D_{\rm max} \ll 1$ proposed by \citet{alard07} 
extended to be applied the tangential critical curve (eq.~\ref{dmax_pnfw}),
is not adequate to set a limit of validity
for these cases. To this end, we use a figure-of-merit,
$\mathcal{D}^2$ (eq. \ref{D2_pm}) to quantify the deviation of the
Perturbative Approach from the exact solution for caustics and critical curves. We
verify that $\mathcal{D}^2$ provides a quantitative description of the
deviation among both solutions. In particular, $\mathcal{D}^2$ decreases
with $\kappa_s$ (as can be drawn from Fig.  \ref{pnfw_ccurves_mp-se}) and
increases with $\varepsilon$ (as expected from the increasing of the perturbation to the lensing
potential with $\varepsilon$). Since the deviation between the exact and perturbative solutions for caustics
is higher than the deviation for critical curves, it is sufficient to set a limit on $\mathcal{D}^2$ for caustics to ensure a small deviation for critical curves.

By setting a threshold on $\mathcal{D}^2$ computed at caustics, a maximum value of $\varepsilon$ is determined for each $\kappa_s$, such that a good matching for caustics and also for critical curves is ensured. We determine these maximum
values  $\varepsilon^{\rm PA}_{\rm max}(\kappa_s)$ by choosing $\mathcal{D}^2=5\times 10^{-4}$. This defines a domain of applicability of the Perturbative Approach for the PNFW model in the range of $\kappa_s$ being considered.
We provide a fitting function for  $\varepsilon^{\rm PA}_{\rm max}(\kappa_s)$ (eq. \ref{emax_pm_fit}).  For $\kappa_s \lesssim 0.8$, the Perturbative Approach is limited to $\varepsilon \lesssim 0.1$. However, for $\kappa_s>1.0$ it is
possible to use this approach even up to $\varepsilon=0.4$ for these cases.

Another limit on the PFNW model parameters is obtained from the comparison of the deformation cross section
for both exact and perturbative calculations.
The fractional deviation is less than 10\% (Fig. \ref{dcs_pnfw-space}) for $\kappa_s \gtrsim 0.7$ and $\varepsilon \lesssim 0.3$ (corresponding to $\epsilon_\Sigma \lesssim 0.55$).

We may use these results to set further constraints on the ellipticity
parameter of the PNFW model, by requiring an agreement with the exact
$\tilde{\sigma}_{R_{\rm th}}$, besides the condition $\varepsilon <
\varepsilon^{\rm PA}_{\rm max}(\kappa_s)$. This ensures that caustics,
critical curves, and the local mapping are well reproduced by the
Perturbative Approach for the PNFW model.  The combined restriction,
imposing the matching for caustics and an agreement to about 10\% for
deformation cross sections, is given in eq. (\ref{region_lt-10}).

In this paper we provided a first systematic attempt to set limits on the domain of applicability of the Perturbative Approach for strong lensing in terms of the parameters of a given lens model, more specifically for the PNFW model. The limits are imposed so that the caustics, critical curves and deformation cross section match the exact solutions with a given accuracy. Although these quantities are useful for strong lensing applications, it is important to determine a domain of validity for arcs/finite sources. For example, \citet{japas} investigated the domain of validity of the
Perturbative Approach for extended circular sources. It is argued that Perturbative Approach can be used for
sources with radius $\lesssim 0.2\,x_{\rm E}$ up to $\varepsilon
\simeq 0.3$. This result should be extended for generic configurations probing the space of the source and lens parameters and their relative position. We expect that the limits here obtained can be connected to the domain of validity for arcs providing guidance to the exploration of this wider parameter space. The systematic application to arcs and connection to the current results is left for a subsequent work.
It is also important to check whether the criterion established here for the  $\mathcal{D}^2$ threshold
can be applied to other lens models, so that we have an {\it a priori} criterium for the domain of validity of the Perturbative Approach regardless of the specific model. 

The usefulness of the Perturbative Approach justifies the search for a determination its accuracy and limit of applicability. Once this is established we will be able to safely use this promising technique in a number of applications, within its domain of validity.

\section*{Acknowledgments}
HSDM is funded by the Brazilian agencies FAPERJ (E-26/101.784/2010), CNPq (PDJ/162989/2011-3), and the PCI/MCTI 
program at CBPF (301.860/2011-4). GBC is funded by CNPq and CAPES. BM is supported by FAPERJ.  
MM is partially supported by CNPq (grant 312876/2009-2) and FAPERJ (grant E-26/110.516/2012).  
MSSG acknowledges the hospitality of the Centro Brasileiro de Pesquisas F\'isicas (CBPF), where part of this work was performed, and the PCI/MCTI program (170.524/2006-0).
We also thank the support of the Laborat\'orio
Interinstitucional de e-Astronomia (LIneA) operated jointly by CBPF, the Laborat\'orio Nacional de
Computa\c c\~ao Cient\'ifica (LNCC), and the Observat\'orio Nacional (ON) and
funded by the Ministry of Science, Technology and Innovation (MCTI). 
MM acknowledges C. Alard for useful discussions regarding the Perturbative Approach for arcs.
We  thank Marcos Lima for useful discussions.

\end{document}